\documentclass[pra,twocolumn,showpacs,amsmath,amssymb,byrevtex]{revtex4}
\usepackage{amsmath,amssymb,amsthm,graphicx,color,bm}

\newtheorem{Thm}{Theorem}

\newcommand{\HA}{\mathop{\mathcal{H}}\nolimits}
\newcommand{\K}{\mathop{\mathcal{K}}\nolimits}

\newcommand{\R}{\mathop{\mathbb{R}}\nolimits}

\newcommand{\I}{I}

\newcommand{\bra}[1]{\langle #1 |}

\newcommand{\ket}[1]{| #1 \rangle}

\newcommand{\bracket}[2]{\langle #1 | #2 \rangle}

\newcommand{\ketbra}[2]{| #1 \rangle \langle #2 |}

\newcommand{\bS}{\mathbf{S}}
\newcommand{\bA}{\mathbf{A}}
\newcommand{\bP}{\mathbf{P}}
\newcommand{\bx}{\mathbf{x}}
\newcommand{\by}{\mathbf{y}}
\newcommand{\cH}{\mathcal{H}}
\newcommand{\cI}{\mathcal{I}}
\newcommand{\cK}{\mathcal{K}}
\newenvironment{Proof}{\begin{trivlist}
\item[\hskip \labelsep {\em \indent Proof.}]}{\qed\end{trivlist}}
\newcommand{\beqa}{\begin{eqnarray}}
\newcommand{\eeqa}{\end{eqnarray}}
\newcommand{\ep}{\epsilon}
\newcommand{\et}{\eta}
\newcommand{\ps}{\psi}
\newcommand{\rh}{\rho}
\newcommand{\De}{\Delta}
\newcommand{\Ga}{\Gamma}
\newcommand{\Tr}{{\rm Tr}}
\newcommand{\braket}[1]{\langle #1 \rangle}
\newcommand{\da}{\dagger}
\newcommand{\cut}[1]{}
\renewcommand{\log}{\ln}

\begin{document}

\title{Quantum Limits of Measurements Induced by Multiplicative Conservation Laws: 
Extension of the Wigner-Araki-Yanase Theorem}

\author{Gen Kimura,$^1$ Bernhard K. Meister,$^2$ and Masanao Ozawa$^3$ }

\affiliation{ $^{1}$Research Center for Information Security, 
National Institute of Advanced Industrial Science and Technology, 
Daibiru Building 1102,
Sotokanda, Chiyoda-ku, Tokyo, 101-0021, Japan,}\email{gen-kimura@aist.go.jp}

\affiliation{ $^2$Physics Department, Renmin University of China,
Beijing, 100872,
 China,}
\email{b.meister@imperial.ac.uk}

\affiliation{ $^{3}$Graduate
School of Information Science, Nagoya University, Chukusa-ku,
Nagoya, 464-8601, Japan.}\email{ozawa@is.nagoya-u.ac.jp}

\begin{abstract}
The Wigner-Araki-Yanase (WAY) theorem shows that additive 
conservation laws limit the accuracy of measurements.
Recently, various quantitative expressions have been found 
for quantum limits on measurements induced 
by additive conservation laws, and have been applied to the study of
fundamental limits on quantum  information processing.  
Here, we investigate generalizations of the WAY theorem to 
multiplicative conservation laws.
The WAY theorem is extended to show that an observable 
not commuting with the modulus of, or equivalently the square of, 
a multiplicatively conserved quantity cannot 
be precisely measured.  
We also obtain a lower bound for the mean-square noise 
of a measurement in the presence of a multiplicatively conserved quantity.
To overcome this noise it is necessary to make large
the coefficient of variation (the so-called relative fluctuation),
instead of the variance as is the case for additive conservation laws,
of the conserved quantity in the apparatus.
\end{abstract}

\maketitle

\section{Introduction}

In recent investigations \cite{02CQC,Lid03,03CQC(R),
03QLM,03UPQ,04UUP,05CQL,06MEP,07CLI} 
it has been established that conservation laws put
a precision limit or compensating resource requirements 
on quantum information processing.
Quantitative analysis suggests that conservation laws lead to 
undesirable entanglement of the object system with the control system, 
such as atom qubits controlled by the Jaynes-Cummings interaction with
an electromagnetic field,
causing decoherence of the object system, even if the 
environment induced decoherence is completely suppressed.
Such limitations would disappear if the control system were considered to be
macroscopic, so that this effect is of a quantum nature of
the control system.

This sort of quantum limit has, however, long been known 
for measurements.
Wigner \cite{Wig52} first claimed  in 1952 that an observable 
which does not commute 
with an additively conserved quantity cannot be measured precisely.
In 1960, Araki and Yanase \cite{AY60} rigorously proved
the impossibility of nondestructive \footnote{
Here, a measurement is ``nondestructive'' if the measured observable is not
disturbed by the interaction between the system and the apparatus.
 Note also that ``a quantum nondemolition (QND) measurement'' is reserved to mean 
a sequence of precise measurements such that the result of each measurement 
is completely predictable from the result of the preceding measurement
 (cf.~p.~363 of  Ref.~\cite{CTDSZ80}). 
Thus, a sequence of nondestructive and precise measurements of
a constant of motion is a QND measurement.}
and precise measurement of  a
discrete observable not commuting with a bounded additively 
conserved quantity such as angular momentum;
the result has been called the Wigner-Araki-Yanase (WAY) theorem.
Subsequently, Yanase \cite{Yan61} found a bound for the accuracy
of spin measurement and concluded that in order to increase 
the accuracy one needs to use a very large measuring apparatus;
see also Wigner \cite{WIg63} and 
Ghirardi, Miglietta, Rimini, and Weber \cite{GMRW81a,GMRW81b}.

In order to extend the WAY theorem to continuous observables,
one of the present authors \cite{91CP,93WA} introduced a quantitative 
approach using commutation relations of noise operators.
He showed the impossibility of nondestructive and precise measurement
of an observable not commuting with additively conserved quantity
in an appropriate limit sense, and yet showed the possibility of arbitrarily precise,
nondestructive measurement of any observable having a c-number commutator 
with additively conserved quantity, such as position measurement under the momentum  
conservation law; see also Stein and Shimony
\cite{SS71}.  
As a quantitative generalization of the WAY theorem, 
a lower bound for the sum of the mean-square noise 
and the mean-square disturbance was obtained in Ref.~\cite{03QLM} 
and used to give a precision limit to realizing universal quantum gates 
in Ref.~\cite{02CQC}. 

There has been a debate as to whether the WAY theorem can be
extended to destructive measurements. 
Ohira and Pearle \cite{OP88} constructed a model by which 
they claimed the possibility of precise measurement of a spin 
component under the spin conservation law in any direction.
However, one of the present authors \cite{02CLU} pointed
out that the claim is a circular argument, since the model
only transfers the problem of measurement of object spin
in one direction to the measurement of the spin of the
probe in the same direction.
In the same paper, it was shown by establishing a  lower bound for the
mean-square noise of the measurement that there is no measurement 
model for the precise measurement of an observable not commuting with a bounded
conserved quantity if the meter observable is required to be
measured nondestructively.

This lower bound was subsequently shown \cite{03UPQ}
to be a consequence of the noise-disturbance uncertainty principle 
\cite{03UVR,04URN},
to solve a long-standing question as to how
the WAY theorem relates to the uncertainty principle,
and was used to derive the limit of the
achievable gate fidelity of realizations of the Hadamard gate 
under the angular momentum conservation law; see also 
Refs.~\cite{05CQL,06MEP,07CLI}.

In this paper, we investigate an entirely new generalization of the 
WAY theorem and show that the WAY theorem can be extended
in a relatively similar formulation to multiplicative conservation laws
qualitatively and quantitatively
\footnote{A preliminary consideration on this topic has already been 
presented in Ref.~\cite{Mei07}.}.
The inevitable mean-square noise induced by the presence of a
multiplicatively conserved quantity is evaluated to show that to overcome
the limit we need to make large 
the coefficient of variation (or the so-called relative fluctuation) of the conserved
quantity in the apparatus. 
We also show that the extension of the WAY theorem to the 
multiplicative case includes the additive case as a corollary. 
Furthermore,  we obtain a limitation on such a measurement of an observable
that the measuring interaction has an invariant state.

In Section \ref{se:II} the concept of generalized measurements is reviewed,
and the measurement models considered by Araki and
Yanase \cite{AY60} are characterized as those measurement models
with both  zero mean-square noise
and zero mean-square disturbance.
In Section \ref{se:III} an extension of the WAY 
theorem to multiplicatively conserved
quantities is proved and an application to quantum statistical
mechanics is discussed. 
In Section \ref{se:IV}  a lower bound for the mean-square noise induced by the
presence of a  multiplicatively conserved quantity is obtained.
Furthermore, an extension of the WAY theorem is also derived 
to destructive measurements under multiplicative conservation laws.

\section{Measuring Processes}
\label{se:II}

A mathematical proof of the Wigner-Araki-Yanase theorem 
was first given by Araki and Yanase \cite{AY60}.
In their formulation, a measuring process is modelled  as
follows. 
Let $A$ be an observable of a quantum system $\bS$ represented 
by a Hilbert
space $\HA$; in this paper, all Hilbert spaces are assumed separable,
or of at most countably infinite dimension.
It is assumed that $A$ has eigenvalues $\mu$ with
corresponding  complete orthonormal eigenvectors $\ket{\phi_{\mu \rho}}$ in
$\HA$, i.e.,  
\beqa\label{eq:observable}
A \ket{\phi_{\mu \rho}} = \mu \ket{\phi_{\mu \rho}}
\eeqa
and  
$\bracket{\phi_{\mu \rho}}{\phi_{\mu^\prime \rho^\prime}} = \delta_{\mu
\mu^\prime} \delta_{\rho \rho^\prime}$.  Here the index $\rho$ represents the
degeneracy parameter of the eigenvalues of $A$.  Following von Neumann \cite{vN55},
it is assumed that  the measurement of $A$ in a state $\ket{\phi} \in \HA$ is
carried out by an interaction with a probe system $\bP$ described by a
Hilbert space $\K$ in its initial state $\ket{\xi}\in\cK$.
The time evolution of the composite system $\bS+\bP$ during
the interaction is represented by a unitary operator $U$ on $\HA\otimes\K$.  
It is assumed that the relation
\begin{equation}\label{eq:U}
U \ket{\phi_{\mu \rho}\otimes \xi} 
= \sum_{\rho^\prime} \ket{\phi_{\mu \rho^\prime}\otimes X_{\mu
\rho\rho^\prime}}
\end{equation}
holds 
with the distinguishability condition
(see footnote 3 of Ref.~\cite{AY60})
\begin{equation}\label{eq:dis}
\bracket{X_{\mu \rho\rho^{\prime}}}{X_{\nu \sigma\sigma^{\prime}}} = 0 \ 
\end{equation}
if $\mu \neq \nu$. 
After the interaction, a meter observable $M$ on $\K$ is measured
to obtain the outcome of the measurement, where $M$ is given by 
\beqa\label{eq:meter}
 M\ket{X_{\mu\rho \rho^\prime}}=\mu \ket{X_{\mu\rho \rho^\prime}}
\eeqa 
for all $\mu,\rho,\rho^\prime$, where $\mu$ varies over a countable
set of real numbers, and $\rho$ and $\rho^\prime$ vary over a countable
index set depending on the value of $\mu$.

The above model describes a class of physically realizable
measurements \cite{vN55}, but is not sufficiently general to include all the
physically realizable measurements. In the modern approach
\cite{84QC,00MN,01OD,04URN},  an exhaustive class of measurements
is formulated as follows. Let $\bA({\bf x})$ be a measuring
apparatus with  macroscopic output variable ${\bf x}$ to measure 
observable $A$  possibly with some error. 
The probe system $\bP$ with a Hilbert space $\K$, a part of the
apparatus, is initially prepared in a state $\ket{\xi} \in \K$, and
interacts with the system $\bS$ during a finite but short time interval
\footnote{Here, we consider instantaneous measurements \cite{01OD} 
in which
the measuring interaction takes place in a time interval so short
that the system time evolution is negligible.  
This is the case if the coupling between
the system and probe is considered to be very strong.},
in which
the composite system $\bS+\bP$ undergoes the time evolution
described by a unitary operator $U$ on $\HA\otimes \K$.  
After the interaction, a meter observable $M$ on $\K$ is measured
to obtain the macroscopic output $\bf{x}$.
According to the Born statistical formula, if the system $\bS$ is initially 
in a state $\ket{\phi} \in\HA$,  the probability distribution of 
the output ${\bf x}$ is given by 
\begin{equation}\label{eq:OPD}
\mathrm{Pr}\{{\bf x} \in \Delta \} = \|[I_1\otimes E^M(\Delta)] U
\ket{\psi \otimes \xi} \|^2,
\end{equation} 
where $E^M(\Delta)$ is the spectral projection of
$M$  corresponding to a Borel set or an interval $\Delta$;
throughout this paper, index 1 refers to the system $\bS$ and 
2 to the probe $\bP$, and accordingly $I_1$ and $I_2$ refer to the
identity operators of $\bS$ and $\bP$, respectively.
The mapping $E^{M}:\De\mapsto E^M(\Delta)$ is called 
the spectral measure of $M$ \cite{Hal51}.

\cut{All the statistical properties of the apparatus
$\bA(\bx)$ is specified by the measuring process $(\cK,\ket{\xi}, U,M)$
as follows.
For any density operator $\rh$ on $\cH$ and Borel set $\De$, 
let $\cI(\De)\rh$ be the operator on
$\cH$ defined by
\beqa
\cI(\De)\rh=\Tr_{\cK}
\{[I_1\otimes E^{M}(\De)]U(\rho\otimes\ketbra{\xi}{\xi})U^{\da}]\}.
\eeqa
Then, the mapping $\cI(\De):\rh\mapsto\cI(\De)\rh$ is a completely positive 
map on the space of trace class operators on $\cH$, called the operation
for the Borel set $\De$ determined by the apparatus $\bA(\bx)$.
The mapping $\cI:\De\mapsto\cI(\De)$ is a completely positive map-valued 
measure, called the instrument determined by the apparatus $\bA(\bx)$.}

It is well-known that every measurement is associated with a probability 
operator-valued measure (POVM) \cite{Hel76}
that describes the output probability
distribution.  The POVM $\Pi$ of the apparatus $\bA(\bx)$ is given by
\beqa\label{eq:POVM}
\Pi(\De)=\braket{\xi|U^{\da}[I_1\otimes E^{M}(\De)]U|\xi}
\eeqa
for all Borel set $\De$,
where $\braket{\xi|\cdots|\xi}$ stands for the partial inner product on
$\cK$; the mapping $\Pi:\De\mapsto\Pi(\De)$ is a positive 
operator-valued measure satisfying the normalization condition $\Pi(\R)=I_1$.
Then, from Eq.~\eqref{eq:OPD} the output probability distribution 
satisfies
\beqa
\mathrm{Pr}\{{\bf x} \in \Delta \}=\|\Pi(\De)^{1/2}\ket{\ps}\|^2.
\eeqa

We say
that the apparatus $\bA(\bx)$  described by the measuring process
$(\cK,\ket{\xi}, U,M)$ {\em precisely measures} observable $A$ if 
the output probability distribution satisfies the
Born statistical formula for observable $A$, i.e., 
\begin{equation}
\mathrm{Pr}\{{\bf x} \in \Delta \} = \|E^A(\Delta) \ket{\psi}\|^2
\end{equation} 
for any state $\ket{\psi} \in \HA$.
In terms of POVM, from \eqref{eq:POVM} this condition is equivalent to
the condition that the POVM of the apparatus coincides with spectral measure
of the observable $A$, i.e., $\Pi=E^{A}$. 
This condition is also equivalent to the relation
\beqa\label{eq:precise2}
U^\dagger [\I_1 \otimes E^{M}(\De)] U\ket{\psi\otimes\xi}
= E^{A}(\De) \otimes \I_2\ket{\psi\otimes\xi}
\eeqa
for any Borel set $\De$ and state $\ket{\ps}\in\cH$,
since two projection operators $U^\dagger (\I_1 \otimes E^{M}(\De)) U$
and $E^{A}(\De) \otimes \I_2$ are identical on the space $\cH\otimes \ket{\xi}$
if and only if their expectation values are identical for all states in that space.
The last condition is also equivalent to the condition 
that in the  Heisenberg picture the meter observable
$M$ precisely evolves to the observable $A$ to be measured, i.e., 
\begin{equation}\label{eq:precise}
U^\dagger (\I_1 \otimes M) U\ket{\psi\otimes\xi}
= A \otimes \I_2\ket{\psi\otimes\xi}
\end{equation}
for any states $\ket{\psi}$ in the domain of $A$,
since two operators $U^\dagger (\I_1 \otimes M) U$ and 
$A$ are identical on the space $\cH\otimes \ket{\xi}$ if and
only if their spectral measures are identical on that space.
To quantify the difference between the both sides,
we introduce the {\it root-mean-square noise} 
$\epsilon(A,\ket{\psi})$ of 
the measurement of $A$ in the state $\ket{\psi}$ 
defined by 
\beqa
\epsilon(A,\ket{\psi})
 = \|N \ket{\psi \otimes \xi}\|,
\eeqa
where the {\it noise operator} $N$ is defined by
\beqa
N=U^\dagger (\I_1 \otimes M) U - A \otimes \I_2.
\eeqa
Then, the precise measurement is equivalently 
characterized by the condition
\begin{equation}
\epsilon(A,\ket{\psi}) = 0 
\end{equation}
for any states $\ket{\psi}$ in the domain of $A$  \cite{04URN}.  

We say that the apparatus $\bA(\bx)$ {\em does not disturb}
an observable $B$ on $\cH$ if the time evolution $U$ does not change 
the probability distribution of observable $B$, i.e.,
\begin{equation}\label{eq:nondisturbance}
\| E^B(\Delta) \ket{\psi}\|^2 = \|[E^B(\Delta)\otimes \I_2] U
\ket{\psi\otimes \xi}\|^2
\end{equation}
for all states $\ket{\psi}\in \HA$. 
This condition is also equivalent to the relation
\beqa\label{eq:disturbance2}
U^\dagger [E^B(\Delta)\otimes \I_2] U\ket{\psi\otimes\xi}
= E^B(\Delta)  \otimes \I_2\ket{\psi\otimes\xi}
\eeqa
for any Borel set $\De$ and state $\ket{\ps}\in\cH$ by a similar
reasoning as above.
The last condition is also equivalent to the condition 
that in the Heisenberg picture
the time evolution during the measuring interaction does not change
the observable $B$, i.e., 
\begin{equation}\label{eq:disturbance}
U^\dagger (B \otimes \I_2) U\ket{\psi\otimes\xi}
= B \otimes \I_2 \ket{\psi\otimes \xi}
\end{equation}
for any states $\ket{\psi}$ in the domain of $B$ \cite{04URN}, or 
the condition that
$B\otimes I_2$ commutes with $U$, i.e., 
\beqa
[B\otimes I_2,U]\ket{\psi\otimes \xi}=0
\eeqa
for any states $\ket{\psi}$ in the domain of $B$  \cite{04URN}.

To quantify the difference between both sides of 
Eq.~\eqref{eq:disturbance}
the {\em root-mean-square disturbance} $\eta(B,\ket{\psi})$
of the observable $B$ in a state $\ket{\psi}$ is naturally defined by
\beqa
\eta(B,\ket{\psi}) =\|D\ket{\psi \otimes \xi}\|,
\eeqa
where the {\it disturbance operator} $D$ is defined by
\beqa
D=U^\dagger (B \otimes\I_2) U - B \otimes \I_2.
\eeqa

According to Eq.~(\ref{eq:disturbance}), the apparatus $\bA(\bx)$ 
does not disturb the observable $B$ if and only if 
$$
\eta(B,\ket{\psi}) = 0 
$$
for all states $\ket{\psi}$  in the domain of $B$   \cite{04URN}.

We say that an  apparatus $\bA(\bx)$  {\em nondestructively and
precisely measures} an observable $A$ if $\bA(\bx)$ precisely measures
$A$ without disturbing $A$, i.e., 
\beqa\label{eq:ndpr}
\ep(A,\ps)=\et(A,\ps)=0
\eeqa
for any state $\ket{\ps}$ in the domain of $A$.

Now, we shall show that the measurement models considered by
Araki and Yanase \cite{AY60} mentioned above are characterized 
by the above condition \eqref{eq:ndpr}.
To formulate the statement, an apparatus $\bA(\bx)$
described by $(\cK,\ket{\xi},U,M)$ is said to be of the {\em Araki-Yanase type}
if there is a complete orthonormal basis $\{\ket{\phi_{\mu \rho}}\}$ in
$\HA$ and a family $\{\ket{X_{\mu\rho \rho^\prime}}\}$ of vectors in $\cK$
satisfying Eqs.~\eqref{eq:U}, \eqref{eq:dis}, and \eqref{eq:meter}.

\begin{Thm}
An apparatus $\bA(\bx)$ described by $(\cK,\ket{\xi},U,M)$ nondestructively
and precisely measures an observable $A$ on $\cH$ if and only if 
$\bA(\bx)$ is of the Araki-Yanase type.  In this case, the observable $A$
is uniquely determined by relation \eqref{eq:observable}.
\end{Thm}
\begin{Proof}
First, suppose that $\bA(\bx)$ is of the Araki-Yanase type
and let $A$ be defined by \eqref{eq:observable}.
Then, from conditions \eqref{eq:U} and \eqref{eq:meter}, we have 
\beqa
U^\dagger (\I_1\otimes M) U \ket{\phi_{\mu \rho} \otimes \xi} 
= A \otimes \I_2 \ket{\phi_{\mu \rho} \otimes \xi}
\eeqa
for any eigenvector $\ket{\phi_{\mu \rho}}$ of $A$.
From the completeness of $\ket{\phi_{\mu \rho}}$ in
$\HA$, the preciseness condition holds, i.e., Eq.~\eqref{eq:precise} holds
for all states $\ket{\psi}$  in the domain of $A$.
Similarly, the condition for nondestructive measurements is shown to be
satisfied.  Next, suppose that $\bA(\bx)$ nondestructively
and precisely measures an observable $A$ on $\cH$.
First, we shall show that the observable $A$ has purely discrete 
spectrum by appealing to a general theorem stating that if the apparatus $\bA(\bx)$
satisfies the repeatability hypothesis, then the observable $A$ has purely 
discrete spectrum (Theorem 6.6 of Ref.~\cite{84QC}).
In order to formulate the repeatability hypothesis, we define the 
joint probability distribution of the repeated measurement using
the apparatus $\bA(\bx)$ by
\beqa\label{eq:joint}
\Pr\{\bx\in\De,\by\in\Ga\}
=
\|[E^{A}(\Ga)\otimes E^{M}(\De)]U\ket{\ps\otimes\xi}\|^2.\nonumber\\
\eeqa
Then, the apparatus $\bA(\bx)$ is said to satisfy the repeatability hypothesis
if
\beqa
\Pr\{\bx\in\De,\by\in\Ga\}=\Pr\{\bx\in\De\cap\Ga\},
\eeqa
or equivalently
\beqa\label{eq:repeatability}
\lefteqn{
[E^{A}(\Ga)\otimes E^{M}(\De)]U\ket{\ps\otimes\xi}
}\quad\nonumber\\
&=&
[I_1\otimes E^{M}(\De\cap\Ga)]U\ket{\ps\otimes\xi}.
\eeqa
Thus, it suffices to show the last relation.
Eq.~\eqref{eq:disturbance2} with $A=B$ implies 
\beqa
[E^{A}(\Ga)\otimes I_2] U\ket{\ps\otimes\xi}
=
U[E^{A}(\Ga)\otimes I_2] \ket{\ps\otimes\xi},
\eeqa
and similarly,  Eq.~\eqref{eq:precise} implies
\beqa
[I_1\otimes E^{M}(\Ga) ]U\ket{\ps\otimes\xi}
=
U[E^{A}(\Ga)\otimes I_2] \ket{\ps\otimes\xi}.
\eeqa
Combining the above two equations, we have
\beqa
\lefteqn{
[E^{A}(\Ga)\otimes I_2]U\ket{\ps\otimes\xi}
}\quad\nonumber\\
&=&
[I_1\otimes E^{M}(\Ga)]U\ket{\ps\otimes\xi}.
\eeqa
Multiplying the both sides by $I_1\otimes E^{M}(\De)$
from left, we obtain Eq.~\eqref{eq:repeatability}.
Thus, the apparatus satisfies the repeatability hypothesis, 
and hence the observable
$A$ has purely discrete spectrum.  Thus, $A$ has 
complete orthonormal eigenvectors $\ket{\phi_{\mu \rho}}$
satisfying \eqref{eq:observable}.
Applying Eqs.~\eqref{eq:precise} and \eqref{eq:disturbance} with $B=A$
to the state $\ket{\psi} = \ket{\phi_{\mu \rho}}$, 
we obtain that $U\ket{\phi_{\mu \rho}\otimes X}$
is an eigenvector belonging to the eigenvalue
$\mu$ of both $A \otimes \I_2$ and $\I_1 \otimes M$. 
Since the eigenspace of
$A \otimes \I_2$ with eigenvalue $\mu$ is spanned by all
$\ket{\phi_{\mu \rho}\otimes X}$ with arbitrary $\rho$ and arbitrary
$\ket{X}\in\K$, vector $U\ket{\phi_{\mu \rho}\otimes X}$ is generally 
written as Eq.~\eqref{eq:U}.
Since this is an eigenvector of $\I_1 \otimes M$, 
the vector $\ket{X_{\mu \rho \rho^\prime}} $ is in the eigenspace of
$\I_1\otimes M $ with eigenvalue $\mu$, and hence satisfies condition
\eqref{eq:dis}. 
Thus, we have shown that $\bA(\bx)$ is of the Araki-Yanase type.
Since the precisely measured observable $A$ is uniquely determined
by the output probability distribution, the uniqueness of $A$ follows
obviously.
\end{Proof}

In the next section, we discuss the limitation on
nondestructive and precise measurements under a multiplicative 
conservation law, while in Section \ref{se:IV} we discuss the limitation
to arbitrary precise measurements with nondestructively 
measurable meters.

\section{Limitation on precise and nondestructive 
measurements induced by multiplicative conservation laws}
\label{se:III}

Let $\bA(\bx)$ be an apparatus described by $(\cK,\ket{\xi},U,M)$
for a Hilbert space $\cH$.
Let $L_1$ and $L_2$ be observables on $\HA$ and $\K$, respectively.  
The observable $L = L_1 \otimes \I_2 + \I_1 \otimes L_2$ is called  an {\em
additively conserved quantity} of $\bA(\bx)$ if $U$ satisfies
\begin{equation}\label{eq:AClaw}
[L_1 \otimes \I_2 + \I_1 \otimes L_2, U] = 0. 
\end{equation}

The additive conservation law is generally associated with a continuous symmetry, 
and often holds for such quantities 
as energy, angular momentum, and
spin.  With reference to the discussions in the preceding section, 
Araki and Yanase \cite{AY60} 
proved

{\bf Wigner-Araki-Yanase Theorem:} {\em An apparatus $\bA(\bx)$ with
an additively conserved 
quantity $L = L_1 \otimes \I_2 + \I_1 \otimes L_2$
nondestructively and precisely measures an observable $A$,
then the observable $A$ must commute with
the conserved quantity, i.e.,  $[A,L_1]  = 0$, 
provided that $L_1$ is bounded.}

An observable $L = L_1 \otimes L_2$ with observables $L_1$
on $\cH$ and $L_2$ on $\cK$ is called  a {\em 
multiplicatively conserved quantity} of $\bA(\bx)$ if $U$ satisfies 
\begin{equation}\label{eq:MClaw}
[L_1 \otimes L_2, U] = 0. 
\end{equation}

Multiplicative conservation laws are related to discrete symmetries 
such as parity, charge conjugation, and time
reversal.  Moreover, they also formally include all the
additive conservation laws by exponentiating the additively
conserved quantities.  
We shall show that a similar limitation to that for the additive case
arises for measurements under 
multiplicative conservation laws.

Let $L_1$ be an observable on $\cH$.
An observable $A$ is said to be {\em nondestructively and
precisely measurable
under the multiplicative conservation law with $L_1$}, if there is an
apparatus $\bA(\bx)$ described by $(\cK,\ket{\xi},U,M)$ such that 
$\bA(\bx)$ precisely measures $A$ without disturbing $A$  and
that $\bA(\bx)$ has a multiplicatively conserved quantity
$L_1 \otimes L_2$ for some invertible observable $L_2$ on $\cK$. 
In the following, $|L|$ stands for the modulus of the
observable $L$ defined by $|L|^2 =L^2$ and $|L|\ge0$.  

\begin{Thm}\label{thm:M-WAY}
Every nondestructively and precisely  measurable observable $A$ 
under the multiplicative conservation law with $L_1$ commutes with
$|L_1|$, i.e., $[A,|L_1|]=0$, provided that  $L_1$ is bounded
and that $L_1$ has a bounded inverse or 0 is an isolated eigenvalue
(if $\cH$ is finite dimensional, the above conditions are automatically
satisfied). 
\end{Thm}
\begin{Proof}
Suppose that $L_1$ is bounded and that $L_1$ has a bounded inverse 
or 0 is an isolated eigenvalue of $L_1$.
Suppose that $A$ can be precisely and nondestructively measured
by an apparatus $\bA(\bx)$ described by $(\cK,\ket{\xi},U,M)$ with
conserved quantity $L_1\otimes L_2$ with an invertible $L_2$.
Let $P_1$ be the projection operator to the kernel of $L_1$. 
Then, the projection operator to the kernel 
of $L_1 \otimes L_2$ on $\HA\otimes\K$
is $P_1 \otimes \I_2$, since $L_2$ is invertible.  From the conservation law
(\ref{eq:MClaw}),  we have $[P_1 \otimes \I_2, U] = 0$ and  
$
[P_1^\perp \otimes \I_2, U] = 0
$.
Then, observing the equality 
$\bra{\phi_{\mu \rho } \otimes \xi} U^\dagger (P_1^\perp \otimes \I_2 )U
\ket{\phi_{\nu \sigma} \otimes \xi} = \bra{\phi_{\mu \rho} \otimes \xi} P_1^\perp
\otimes \I_2 \ket{\phi_{\nu \sigma} \otimes \xi}$  and
Eq.~\eqref{eq:U}, we have 
\begin{eqnarray*}
\lefteqn{\sum_{\rho^\prime,\sigma^\prime} 
\bra{\phi_{\mu \rho^\prime }}P_1^\perp\ket{\phi_{\nu \sigma^\prime}}
\bracket{X_{\mu \rho \rho^\prime}}{X_{\nu \sigma \sigma^\prime}} }\qquad\\
&=&
\bra{\phi_{\mu \rho }}P_1^\perp\ket{\phi_{\nu \sigma}} \bracket{\xi}{\xi} 
\end{eqnarray*}
 for all $\mu,\nu$.
From condition \eqref{eq:dis}, if  $\mu \neq \nu$, we have
\begin{equation}\label{eq:key}
\bra{\phi_{\mu \rho }}P_1^\perp\ket{\phi_{\nu \sigma}} = 0,
\end{equation}
and hence $[A,P_1^\perp]=0$.
Let $\cH'=P_1^\perp\cH$.  
Then, from the above, $\cH'\otimes\cK$ is invariant under $U$,
and $\cH'$ is invariant under $A$ and $L_1$.
Let $A'$ and $L'_1$ be the restrictions of $A$ and $L_1$ to $\cH'$, respectively.  
Let $U'$ be the restriction of $U$ to $\cH'\otimes \cK$.
Then, the measuring process $(\cK,\ket{\xi},U',M)$ precisely and nondestructively 
measures $A'$ with multiplicatively conserved quantity $L'_1\otimes L_2$.
Since $L'_1\otimes L_2$ is invertible, the observable
$\log|L'_1 \otimes L_2|$ is well-defined and satisfies
$$
\log |L'_1 \otimes L_2| = \log|L'_1|\otimes \I_2 + I'_1 \otimes\log|L_2|, 
$$  
where $I'_1$ is the identity on $\cH'$.
Then, we have $[U,\log|L'_1 \otimes L_2|]=0$, so that 
$\log|L'_1|\otimes \I_2 + I'_1 \otimes\log|L_2|$ is an additively conserved 
quantity for $(\cK,\ket{\xi},U',M)$.  It follows from assumptions on $L_1$
that $\log|L'_1|$ is bounded.   Thus, from the Wigner-Araki-Yanase theorem
\cite{AY60} it follows that $[A',\log|L'_1|]=0$.
Thus, we have $[A',|L'_1|]=0$ from $|L'_1|=\exp\log|L'_1|$.
Therefore, we have $[A,|L_1|]=[A,|L_1|P_1^{\perp}]=
[AP_1^{\perp},|L_1|P_1^{\perp}]=0$
\end{Proof}

In comparing the WAY theorem for additive conservation laws
with the present extension to multiplicative conservation
laws, the following two features should be noticed.
First, the measured observable $A$ is required to  commute with $|L_1|$ not
with $L_1$.  Second, for the theorem to be valid, we need an additional assumption
that $L_2$ is invertible.  Indeed, these are necessary assumptions in the multiplicative
cases as the following examples show. Let $\HA$ and $\K$ be two dimensional
Hilbert spaces, and for $i=1,2$ let  $\sigma^{(i)}_x,\sigma^{(i)}_y,\sigma^{(i)}_z$ 
be Pauli matrices for system $\HA$ if $i=1$, and $\K$ if $i=2$, where
$\{\ket{a_i}\}_{i=1,2}$ and $\{\ket{\xi_j}\}_{j=1,2}$ are eigenvectors of
$\sigma^{(1)}_z$ and $\sigma^{(2)}_z$, respectively.  Let us consider a
measurement of $A = \sigma^{(1)}_z$.   With the measuring interaction $U$, 
which
is a controlled-NOT gate defined by $U = \ketbra{a_1}{a_1} \otimes \I_1 +
\ketbra{a_2}{a_2} \otimes \sigma^{(2)}_x$, it is easy to check that the relation
\begin{equation}\label{eq:exU} U \ket{a_i}\otimes \ket{\xi_1} = \ket{a_i}\otimes
\ket{\xi_i} 
\end{equation} holds for $i=1,2$, and hence the apparatus $\bA(\bx)$ described by
$(\cK,\ket{\xi_1},U,M)$ nondestructively and precisely measures 
$A$. 

{\em Example 1}. Let $L_1 = \sigma^{(1)}_x $ and $L_2 = \sigma^{(2)}_x$;
notice that $\sigma^{(2)}_x$ is invertible.  It is easy to check that
$L_1\otimes L_2$ satisfies the multiplicative conservation law with respect to
$U$ in \eqref{eq:exU}.   However, the observable $A = \sigma_z^{(1)}$ does
not commute with $L_1 = \sigma_x^{(1)}$, while $A$ commutes with
$|L_1|= \I_1$.  
This shows that the commutativity with $A$ applies to $|L_1|$ and
not to $L_1$ itself.  
Note that it is natural to impose the invertibility restriction on $L_2$, 
since we have the trivial counterexample $L_2 = 0$. 

{\em Example 2}. Let $L_1$ be an arbitrary observable on $\HA$ and let $L_2$
be a noninvertible observable on $\K$ defined by $L_2 =
\ketbra{\tilde{\xi}}{\tilde{\xi}}$ with $\ket{\tilde{\xi}}
=\frac{1}{\sqrt{2}}(\ket{\xi_1}+\ket{\xi_2})$.  Then, it is easy to see
$[U,L_1\otimes L_2]=0 $.  However there exists an observable $L_1$ with
which $|L_1|$ does not commute with $A$, e.g., $L_1 = |L_1| =
\ketbra{\tilde{a}}{\tilde{a}}$ with $\ket{\tilde{a}}
=\frac{1}{\sqrt{2}}(\ket{a_1}+\ket{a_2})$. 
This shows that the invertibility of $L_2$ 
is a necessary assumption for Theorem \ref{thm:M-WAY} to hold.  

We shall now show that  the usual WAY theorem for additive conservation
laws is obtained as a corollary of Theorem \ref{thm:M-WAY}. 
Let $L$ be an additively conserved quantity for an apparatus $\bA(\bx)$
as defined in \eqref{eq:AClaw} with $L_1$ bounded. 
Then, $\bA(\bx)$ has a multiplicatively conserved quantity $\exp(L_1) \otimes
\exp(L_2)$ satisfying \eqref{eq:MClaw}. 
In this case, $\exp(L_2)$ is invertible and $\exp(L_1)$ is bounded and has 
a bounded inverse, so that
Theorem \ref{thm:M-WAY} concludes $[A,|\exp(L_1)|] = 0$.
Since $\exp(L_1)$ is positive, we have
$[A,\exp(L_1)] = 0$ and hence $[A,L_1] = 0$. 
It is important to point out that Theorem \ref{thm:M-WAY} for multiplicative
conservation laws is not directly obtained from the usual WAY theorem by 
taking the logarithm of the multiplicatively conserved quantity.
Since $L_1$ in condition \eqref{eq:MClaw} is not assumed to be invertible, 
$\log |L_1\otimes L_2|$ is not necessarily an additively conserved quantity.

Another application of Theorem \ref{thm:M-WAY} leads to a limitation on 
a measurement such that the measuring interaction has an invariant state.

\begin{Thm}\label{cor:M-WAYforOB}
Suppose that an apparatus $\bA(\bx)=(\cK,\ket{\xi},U,M)$ 
nondestructively and precisely measures an observable 
$A$ on $\cH$. 
If the measuring interaction $U$ leaves a product state $\rho_1 \otimes \rho_2$
invariant, where $\rho_1$ is a density operator on $\HA$ with finite rank and
$\rho_2$ is an invertible density operator (such as the Gibbs state)
on $\cK$, we have 
\beqa
[A,\rho_1] = 0. 
\eeqa 
\end{Thm}

The above theorem follows easily from 
Theorem  \ref{thm:M-WAY}  with $L_1=\rho_1$ and $L_2=\rho_2$.

\section{Lower bound for the mean-square noise 
and limitation on arbitrary precise measurements}
\label{se:IV}

Now we consider quantitative limitations to measurements under multiplicative
conservation laws. 
For this purpose, a technique previously developed in Ref.~\cite{02CLU}
is used, and we obtain a bound for the  mean-square noise $\epsilon(A,\ket{\psi})$ under
a multiplicative conservation law. 

Suppose that an apparatus $\bA(\bx)$ described by $(\cK,\ket{\xi},U,M)$ has
a multiplicatively conserved quantity $L_1\otimes L_2$.
From the Heisenberg-Robertson's uncertainty relation 
for standard deviations of the noise operator $N$ and the modulus
of the multiplicatively conserved quantity, $|L_1\otimes L_2| =|L_1| \otimes |L_2|$, 
in the state
$\ket{\psi\otimes \xi}$, we have
\beqa
\sigma(N)^2\sigma(|L_1\otimes L_2|)^2 \ge \frac{1}{4}|\langle [N,|L_1|\otimes |L_2|]\rangle|^2,  
\eeqa
where $\sigma(X)$ and $\langle X \rangle$ denote the standard deviation and 
the mean of $X$ in state $\ket{\psi\otimes \xi}$, respectively.  
Since $[U,|L_1|\otimes |L_2|]=0$, we have
the commutation relation
\begin{widetext}
\beqa
[N,|L_1|\otimes |L_2|]=
[A,|L_1|]\otimes L_2-U^{\dagger}(|L_1|\otimes [M,|L_2|])U.
\eeqa
Observing the relations $\epsilon(A,\ket{\psi})^2 \ge 
\epsilon(A,\ket{\psi})^2 - \langle N \rangle^2
 = \sigma(N)^2$ and $\bra{\psi}|L_1|^2 \ket{\psi} \bra{\xi}|L_2|^2 \ket{\xi} =
\langle |L_1\otimes L_2|^2 \rangle \ge \sigma(|L_1\otimes L_2|)^2$,  
we obtain 
\begin{equation}
4\epsilon(A,\ket{\psi})^2\bra{\psi}|L_1|^2 \ket{\psi} \bra{\xi}|L_2|^2 \ket{\xi} 
\ge |\braket {[A,|L_1|]\otimes |L_2| - U^\dagger (|L_1| \otimes [M,|L_2|]) U}|^2.
\label{eq:inequality}
\end{equation}
Let ${\rm Ker(L_j)}$ for $j=1,2$ denote the kernel of $L_j$ in $\cH$ or $\cK$,
respectively.
Let $\ket{\psi}\not\in{\rm Ker}(L_1)$ and $\ket{\xi}\not\in{\rm Ker}(L_2)$.
Then, we have
$\bra{\psi}|L_1|^2 \ket{\psi} \bra{\xi}|L_2|^2 \ket{\xi} \not=0$ and hence
we obtain a bound for the mean-square noise as 
\begin{equation}\label{eq:inequality2}
\epsilon(A,\ket{\psi})^2 
\ge \frac{|\braket {[A,|L_1|]\otimes |L_2| - U^\dagger (|L_1| \otimes [M,|L_2|]) U}|^2
}{4\bra{\psi}|L_1|^2 \ket{\psi} \bra{\xi}|L_2|^2 \ket{\xi}}.
\end{equation}
\end{widetext}

Now, we require Yanase's condition \cite{Yan61,02CLU}
\begin{equation}\label{eq:Yanase}
[M,|L_2| ] = 0.
\end{equation}
This condition eliminates a circular argument to show the measurability 
of $A$ under a conservation law
by reducing it to the measurability of $M$ whose
measurability is still unresolved under the same conservation law.
In order to ensure the measurability of $A$, 
even if we allow the measurement of $A$ to be destructive,
it is natural to assume that there is a measuring process in which 
the meter measurement can be done nondestructively 
to achieve the stability of the measurement outcome to be recorded.
Then, this is possible in the presence of a multiplicative conservation 
law with $L_2$ only if the relation $[M,|L_2|] = 0$ holds.  
Thus, it is natural to require the existence of a measuring process in
which Yanase's condition holds for the meter observable.

Under Yanase's condition we have 
\begin{equation}\label{eq:bound}
\epsilon(A,\ket{\psi})^2 \ge 
\frac{|\bra{\psi} [A,|L_1|]\ket{\psi}| }{4\bra{\psi}|L_1|^2 \ket{\psi}} R(|L_2|),
\end{equation}
where $R(|L_2|)$ is the ratio of the squared mean of $|L_2|$ to the mean 
of $L_2^2=|L_2|^2$ in state $\ket{\xi}$, i.e., 
\beqa
R(|L_2|) \equiv \frac{\bra{\xi}|L_2|\ket{\xi}^2}{\bra{\xi}|L_2|^2\ket{\xi}} \le 1,
\eeqa
where the last inequality holds for any state $\ket{\xi}$ since 
$\bra{\xi}|L_2|^2\ket{\xi} - \bra{\xi}|L_2|\ket{\xi}^2 = \sigma(|L_2|)^2 \ge 0$.  
The ratio $R(|L_2|) $ is directly related to the coefficient of variation
 (relative fluctuation) ${\rm CV}(|L_2|)$ of $|L_2|$, 
the ratio of the standard deviation of $|L_2|$ to the mean of
$|L_2|$ in state $\ket{\xi}$, i.e., 
\beqa
R(|L_2|) &=& 
\frac{\bra{\xi}|L_2|\ket{\xi}^2}{\sigma(|L_2|)^2 + \bra{\xi}|L_2|\ket{\xi}^2} 
\nonumber\\
&=&  \frac{1}{1+{\rm CV}(|L_2|)^2}.
\eeqa

Let $L_1$ be a bounded observable on $\cH$.
An observable $A$ on $\cH$ is said to be 
{\em precisely measurable under the
multiplicative conservation law with $L_1$} if there is an 
apparatus $\bA(\bx)$ described by $(\cK,\ket{\xi},U,M)$ 
such that $\bA(\bx)$ precisely measures $A$, 
that $\bA(\bx)$ has a multiplicatively conserved quantity
$L_1 \otimes L_2$ for some invertible and bounded observable $L_2$
on $\cK$, and that $\bA(\bx)$ satisfies Yanase's condition.
Then, we obtain the following generalization of the WAY theorem.

\begin{Thm}
Every precisely measurable observable
under the multiplicative conservation law with $L_1$
commutes with $|L_1|$.
\end{Thm}
\begin{Proof}
Let $\bA(\bx)$ be an apparatus, described by $(\cK,\ket{\xi},U,M)$, to
carry out a precise measurement of $A$ having a multiplicatively conserved quantity 
$L_1\otimes L_2$, where $L_1$ is bounded and $L_2$ is bounded and invertible,
and satisfying Yanase's condition.
We can assume without any loss of generality that $A$ is bounded; otherwise,
replace $A$ by $\tan^{-1} A$ and $M$ by $\tan^{-1} M$ for instance.
Then, we have $\epsilon(A,\ket{\psi})=0$ for any state $\ket{\psi}\in\cH$.
Let $\ket{\psi}\in\cH$ and $\ket{\xi}\in\cK$.
Then, Eq.~\eqref{eq:inequality} concludes 
$\bra{\ps}[A,|L_1|]\ket{\ps}\bra{\xi}|L_2|\ket{\xi}=0$ under Yanase's  condition.
Since $L_2$ is invertible, we have $\bra{\xi}|L_2|\ket{\xi}>0$, and
$\bra{\ps} [A,|L_1|]\ket{\ps}=0$ holds.  
Therefore, we conclude $[A,|L_1|]=0$.
\end{Proof}

Note that the condition $[A,|L_1|]=0$ is equivalent with $[A,L_1^2]=0$.
Thus, we have
shown that an observable not commuting with the modulus of, or equivalently 
the square of, a multiplicatively conserved quantity cannot 
be precisely measured.  

To figure out the apparatus state $\ket{\xi}$ which makes the measurement
of $A$ as precise as possible, let us consider the state that minimizes 
$R(|L_2|)$; for simplicity we assume $L_2$ to be of finite rank.
If $|L_2|$ is constant, then $R(|L_2|)$ is always
$1$. Suppose that $|L_2|$ is not constant, 
and let $l_m$ and $l_M$ be the minimum and
maximum eigenvalues of $|L_2|$, respectively.
Then, $R(|L_2|)$ takes the
minimum value $\frac{4l_m l_M}{(l_M-l_m)^2} < 1$.
Indeed, it is straightforward to prove the following statement in discrete probability
theory:   Let $L$ be a random variable with values $(l_1,l_2,\ldots,l_d)$ 
with $d\ge 2$ and $0<l_1 < l_2 < \cdots < l_d$.   The ratio 
\beqa
R(L) = \frac{(\sum_{i=1}l_i p_i)^2}{\sum_{i=1}l_i^2 p_i}
\eeqa
with the probability distribution $ (p_1,p_2,\ldots,p_d) $ has the minimum
$\frac{4 l_1 l_d}{(l_d-l_1)^2}$ with the unique probability distribution $
(\frac{l_d}{l_1+l_d},0,\ldots,0,\frac{l_1}{l_1+l_d})$,  while
the variance of $L$ is  maximized by $(\frac{1}{2},0,\ldots,0,\frac{1}{2})$.  The
above minimum is attained by any state $\ket{\xi_{min}}$ with the following
properties:
The probabilities
to obtain the outputs $l_m$ and $l_M$ are $\frac{l_M}{l_m+l_M}$ and
$\frac{l_m}{l_m+l_M}$, respectively, while the probabilities to obtain the other
eigenvalues are zero.
Therefore, any such state $\ket{\xi_{min}}$ can be written as $\ket{\xi_{min}} =
\sqrt{\frac{l_M}{l_m+l_M}}\ket{m} + \sqrt{\frac{l_M}{l_m+l_M}}
\ket{M}$ with $\ket{m}$ and $\ket{M}$ being eigenstates of $|L_2|$ 
with eigenvalues $l_m$ and $l_M$, respectively.  

\section{Concluding remarks}

In this paper natural generalizations of the WAY theorem to multiplicatively conserved
quantities have been established.  We have characterized nondestructive and precise
measurements of an observable by the root-mean-square noise and disturbance.
We have proved that every nondestructively and precisely measurable observable
under the multiplicative conservation law with arbitrary $L_1$ and invertible $L_2$ commutes 
with $|L_1|$; here, we confine ourselves to the finite dimensional case for simplicity.
By taking exponentials, every
additive conservation law can be regarded as a multiplicative conservation law,
such that both $L_1$ and $L_2$ are invertible.  
Thus, the original WAY theorem is recovered as a simple corollary.  
An example shows that there is a nondestructively and precisely measurable observable 
not commuting with $L_1$ under the multiplicative conservation law with arbitrary
$L_1$ and invertible $L_2$.  
Thus, the noncommutativity applies to $|L_1|$ instead of $L_1$.
Another example shows that the invertibility of $L_2$ cannot be dropped from
the assumptions of the above statement.
An interesting application of the above statement is given to invariant states of 
measuring interactions.

We have also investigated destructive measurements 
to drop the assumption of measurements
to be nondestructive from the above statement.  In this case,
a general lower bound for the root-mean-square noise is established in 
Eq.~\eqref{eq:bound} under Yanase's condition that ensures that the 
meter observable can be precisely and nondestructively measured 
with the same conserved quantity. 
This condition is necessary for eliminating a circular argument.
The above lower bound shows that in order to overcome the limitation
we need to make large the coefficient of variation of the conserved
quantity in the apparatus. 
Then, we have concluded that every precisely measurable observable
under the multiplicative conservation law with bounded $L_1$ and 
bounded and invertible $L_2$ commutes with $|L_1|$.

An interesting problem in experimental settings is to obtain 
a tighter lower limit for the root-mean-square noise in the presence of a
multiplicatively conserved quantity; our result is the most general one 
but more specific results can be tighter than the most general.
Theorems in the present paper are stepping stones towards 
developing a WAY type theorem for
conserved quantities that has both additive as well as multiplicative 
components, e.g. a
Hamiltonian of a multiparticle system with an interaction term. 

A broad area of applications of the present investigation is general 
implementation limitations on quantum computers. 
For additive conservation laws, there has been an extensive literature
\cite{02CQC,Lid03,03CQC(R),
03QLM,03UPQ,04UUP,05CQL,06MEP,07CLI} 
on the conservation-law-induced quantum limits on the performance of
elementary quantum gates; see also \cite{BW99,Ban02,EK02,Ita03,EK03,Ban03,
SD04} for the model-dependent approach to the limitations of quantum gate
operations realized by the atom-field interaction, which has turned out to be 
consistent with the model-independent approach based on conservation laws. 
Our method will be also expected to contribute to the problem 
of programmable quantum processors
\cite{NC97,VC00,HZB06} and related subjects 
\cite{DP05,DP05b,DP05c} in future investigations.

\begin{acknowledgments}
This work was supported in part by the SCOPE project of the MIC,
the Grant-in-Aid for Scientific Research (B) Grant No.~17340021 of the JSPS,
and the CREST project of the JST.
G. K. was supported by the Grant-in-Aid for JSPS Research Fellows.
B.K.M. gratefully acknowledges the financial
assistance of the NSF of China.
\end{acknowledgments}


\end{document}